\documentclass[twocolumn,showpacs,aps,prl]{revtex4}
\usepackage{graphicx}
\usepackage{dcolumn}
\usepackage{bm}
\newcommand{\todaye}{\the\day/\the\month/\the\year}

\begin{document}

\preprint{}

\title{Quantum oscillation of magnetoresistance in tunneling junctions \\
       with a nonmagnetic spacer}

\author{H. Itoh$^a$, J. Inoue$^{bc}$ A. Umerski$^d$, and  J. Mathon$^e$}

\affiliation{
$^a$Department of Quantum Engineering, Nagoya University, 
    Nagoya 464-8603, Japan \\
$^b$Department of Applied Physics, Nagoya University, 
    Nagoya 464-8603, Japan \\
$^c$CREST, JST (Japan Science and Technology), 
    Shibuya-ku, Tokyo 150-0002, Japan \\
$^d$Department of Applied Mathematics, Open University, 
    Milton Keynes, MK7 6AA, U.K. \\
$^e$Department of Mathematics, City University, 
    London EC1V 0HB, U.K. }

\date{02/05/2002}

\begin{abstract}
We make a theoretical study of the quantum oscillations of the tunneling 
magnetoresistance (TMR) as a function of the spacer layer thickness. 
Such oscillations were recently observed in tunneling junctions 
with a nonmagnetic metallic spacer at the barrier-electrode interface. 
It is shown that momentum selection due to the insulating barrier 
and  conduction via quantum well states in the spacer, mediated by diffusive 
scattering caused by disorder, are essential features required 
to explain the observed period of oscillation in the TMR ratio 
and its asymptotic value for thick nonmagnetic spacer. 
\end{abstract}

\maketitle


Large magnetoresistance \cite{miyazaki,moodera1} observed 
in ferromagnetic tunneling junctions 
such as Fe/Al$_{2}$O$_{3}$/Fe and Co/Al$_{2}$O$_{3}$/CoFe 
currently attracts much interest due to the possibility of its application 
to magnetic sensors and MRAM elements. 
Because the tunneling magnetoresistance (TMR) ratio 
is related to the spin polarization 
of the ferromagnetic leads \cite{julliere,maekawa}, 
attempts have been made to fabricate junctions with more highly 
spin-polarized ferromagnets \cite{lu,seneor}. 
Realistic calculations \cite{maclaren,mathon3}, on the other hand, have given 
much higher TMR ratios than the observed values, 
which is probably due to their assumption of epitaxial structures. 
Recent experiments 
on TMR using epitaxial junctions \cite{yuasa1,wulfhekel,bowen}, 
however, were unsuccessful in producing TMR ratios as high as expected. 
Thus, our understanding of the relationship between the electronic structure 
of the ferromagnets and the TMR ratio is far from complete.

The most important factor governing the TMR ratio 
may be the electronic structure at junction interfaces \cite{teresa,oleinik}. 
In order to clarify its role,
several experiments have been performed to measure the dependence
of TMR ratio on  the thickness of a nonmagnetic metal layer
inserted at the interface \cite{moodera2,sun,leclair}. 
The observed TMR ratios show almost monotonic decrease 
with increasing thicknesses of inserted layers of Au, Cu, or Cr, 
contrary to a theoretical study for clean junctions \cite{mathon2} 
which shows clear oscillations of the TMR ratio 
as a function of the nonmagnetic layer thickness. 
Zhang and Levy \cite{zhang} have successfully explained 
this decrease in TMR ratio in terms of the decoherence of electron 
propagation across a nonmagnetic layer. 
However, recent experiments by Yuasa et al. 
show clear oscillations of the TMR ratio  as a function of 
Cu layer thickness for  high quality 
NiFe/Al$_{2}$O$_{3}$/Cu/Co junctions in which 
the Co/Cu electrode is a single crystal \cite{yuasa2}. 
In their experiments, two characteristic features of the oscillations 
have been observed: (i) the average TMR ratio decays to zero 
with increasing nonmagnetic layer thickness; 
(ii) the period of the oscillations 
is  determined solely by the belly or long period Fermi wave vector $k_{\mathrm{F}}$ of Cu. 
The observed period agrees quite well 
with that of the oscillations of photoemission spectra 
caused by quantum well states in Co/Cu multilayers \cite{ortega}. 
From the theoretical point of view this is confusing, since in addition to the Fermi wave vector \cite{vedyayev}, 
another wave vector, i.e., the cut-off k-point $k_{\mathrm{cp}}$, 
given by the depth of the quantum well, is also known to contribute 
to the conductance oscillations \cite{mathon1}. 
This wave vector dominates the  predicted oscillations of CPP-GMR 
in a Co/Cu/Co trilayer \cite{mathon4}. 
In fact, the calculated oscillations of TMR 
for a clean junction \cite{mathon2} 
cannot be explained by a single period determined by $k_{\mathrm{F}}$ only. 
Furthermore, the asymptotic value of the TMR ratio calculated 
for a thick spacer layer is finite, 
which disagrees with the observed results. 
The purpose of the present work is to reconcile the theoretical results 
with the observed ones and thus 
deepen our understanding of the TMR effect.

In this Letter we will show that the combined effects of barrier thickness 
and disorder can explain the experimental results. 
In particular, we will demonstrate that 
i) increasing  barrier thickness increases the amplitude of 
the $k_{\mathrm{F}}$ oscillation period 
relative to the $k_{\mathrm{cp}}$ oscillation period, 
ii) disorder introduced in the barrier also weakens 
the amplitude of the $k_{\mathrm{cp}}$  oscillation period, 
and iii) the disorder decreases the asymptotic value of the TMR ratio. 
These results are interpreted in terms of the momentum selection of electrons 
incident on the barrier interface 
and in terms of the diffusive scattering due to disorder which 
opens additional conduction channels via quantum well states. 
In the first part of this Letter, these effects will be demonstrated by 
numerical calculation for a single-orbital tight-binding model. 
We will then demonstrate that the calculated results can be reproduced 
by the stationary phase approximation \cite{mathon4}. 
This implies that this technique is applicable to a realistic multi-orbital 
tunneling junction, where a purely numerical calculation would be unfeasible. 
The results indicate that $k_{\mathrm{F}}$ of Cu is really responsible 
for the oscillation period observed. 

Let us consider a FM/I/NM/FM junction on a simple cubic lattice with 
lattice spacing $a$, where FM, I, and NM denote a ferromagnetic electrode, 
an insulating barrier, and a nonmagnetic metallic spacer, respectively. 
Initially we adopt a single-orbital tight-binding Hamiltonian 
in order to model a Co/Al$_{2}$O$_{3}$/Cu/Co junction: 
\begin{equation}
H=-t\sum_{\left( {i,j}\right) ,\sigma } 
c_{i\sigma }^{\dag }c_{j\sigma }^{\phantom{\dag}} 
+ \sum_{i,\sigma } V_{i\sigma } 
c_{i\sigma }^{\dag }c_{i\sigma}^{\phantom{\dag}}\,, 
\end{equation}
where $c_{i\sigma }^{\phantom{\dag}}(c_{i\sigma }^{\dag })$ is 
the annihilation (creation) operator of an electron with spin $\sigma $ 
at site $i$, $t$ the hopping integral between nearest neighbor sites, 
and $V_{i\sigma}$ the on-site potential for an electron with spin $\sigma $ 
at site $i$. 
Since the majority ($+$) spin band of Co is similar to the Cu band, 
we assume that $V_{\mathrm{FM}+}$ is equal to $V_{\mathrm{NM}}$. 
Figure 1 shows the potential profile of the system. 
Quantum well states are formed in NM only for electrons 
with minority ($-$) spin in the right FM. 
Since the insulating Al$_{2}$O$_{3}$ barrier is amorphous in real junctions, 
we introduce disorder in the barrier 
by requiring that $V_{i\sigma }$ takes $V_{\mathrm{I}}+\Delta V$ 
or $V_{\mathrm{I}}-\Delta V$ values randomly depending on the site 
in the barrier. 

\begin{figure}[tb]
\begin{center}
\leavevmode
\includegraphics[width=0.75\linewidth,clip]{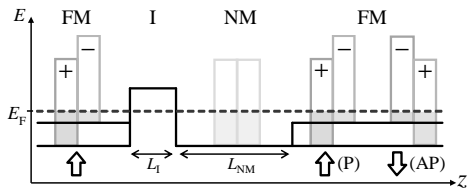}
\caption{Potential profile of a FM/I/NM/FM junction. }
\label{fig1}
\end{center}
\end{figure}

The Kubo formula and a recursive Green's function method 
are used to calculate the tunneling conductances 
$G_{++}$, $G_{--}$, $G_{+-}$, and $G_{-+}$, where $G_{++}$ and $G_{--}$ are 
the conductances in parallel alignment for $\uparrow $ and $\downarrow $-spin 
electrons, respectively, 
and $G_{+-}$ and $G_{-+}$ are those in antiparallel alignment 
for $\uparrow $ and $\downarrow $-spin electrons, respectively. 
The conductance is given by
\begin{equation}
G_{\sigma \sigma ^{\prime }} = \frac{e^{2}}{h}
\sum_{\mathbf{k}_{\parallel },\mathbf{k^{\prime }}_{\parallel }}
t_{\sigma \sigma ^{\prime }}
(\mathbf{k}_{\parallel }\rightarrow \mathbf{k^{\prime }}_{\parallel }) \,,
\end{equation}%
where $t_{\sigma \sigma ^{\prime }}
(\mathbf{k}_{\parallel }\rightarrow \mathbf{k^{\prime }}_{\parallel })$ 
is the transmission coefficient for an electron incident from the left FM 
with $\mathbf{k}_{\parallel }$ and scattered to the right FM 
with $\mathbf{k^{\prime }}_{\parallel }$. 
TMR ratio is evaluated from the conductances 
in the parallel and antiparallel alignments 
as $TMR\equiv 1-(G_{+-}+G_{-+})/(G_{++}+G_{--})$. 
In order to treat the disorder introduced in the insulating barrier, 
we use the single-site coherent potential approximation (CPA). 
The vertex correction to the conductance, 
which describes diffusive scattering, 
is calculated consistently with the coherent potential (self-energy) 
so that the current conservation is satisfied \cite{itoh2}. 
We have also performed numerical simulations \cite{itoh1} 
for finite-size clusters and checked that the results obtained 
by the two methods agree. 

In these numerical calculations, 
we use $V_{\mathrm{FM}+}=V_{\mathrm{NM}}=2.382t$, 
$V_{\mathrm{FM}-}=5.382t$, and Fermi energy $E_{\mathrm{F}}=0.0$. 
The choice of these parameters gives commensurate periods of oscillation 
as shown below. 
As for the insulating barrier, parameters $V_{\mathrm{I}}=9.0t$ 
and $\Delta V=0$ are used for clean junctions, 
and $V_{\mathrm{I}}=9.0t$ and $\Delta V=0.5t$ are used 
for disordered junctions. 
We only show the calculated results for disorder 
within the insulating barrier. 
However, we have checked that the results are not changed 
qualitatively even when we introduce disorder at the interface 
between the nonmagnetic spacer and the ferromagnetic electrode. 

Figures 2(a) and 2(b) show, respectively, the spin-dependent conductances 
and TMR ratios of junctions without disorder. 
It can be seen that $G_{--}$ and $G_{+-}$ oscillate 
with the NM layer thickness $L_{\mathrm{NM}}$ 
due to interference effects caused by the quantum well. 
These oscillations show more than one  period. 
The Fermi wave vector $k_{\mathrm{F}}$ and 
the cut-off k-point $k_{\mathrm{cp}}$ of NM are given 
by $2t\cos (k_{\mathrm{F}}a)=V_{\mathrm{NM}}-E_{\mathrm{F}}-4t$ 
and $2t\cos (k_{\mathrm{cp}}a)=V_{\mathrm{FM}-}-V_{\mathrm{NM}}$, 
respectively \cite{mathon1}. 
Therefore, the periods of oscillation estimated from 
$k_{\mathrm{F}}=4\pi /5a$ and $k_{\mathrm{cp}}=2\pi /3a$ 
are $5a$ and $3a$, respectively. 
The periods of oscillation in $G_{--}$ and $G_{+-}$ shown in Fig.~2(a) 
may be interpreted as a superposition of these two periods 
as discussed later. 
The situation is analogous to that of CPP-GMR 
in a Co/Cu/Co trilayer \cite{mathon4}. 
The TMR ratio shown in Fig.~2(b) oscillates with the same periods as the 
conductance and has a finite asymptotic value for large NM thicknesses. 
These results are consistent with the previous results \cite{mathon2} 
where the effect of disorder was ignored. 

\begin{figure}
\begin{center}
\leavevmode
\includegraphics[width=0.70\linewidth,clip]{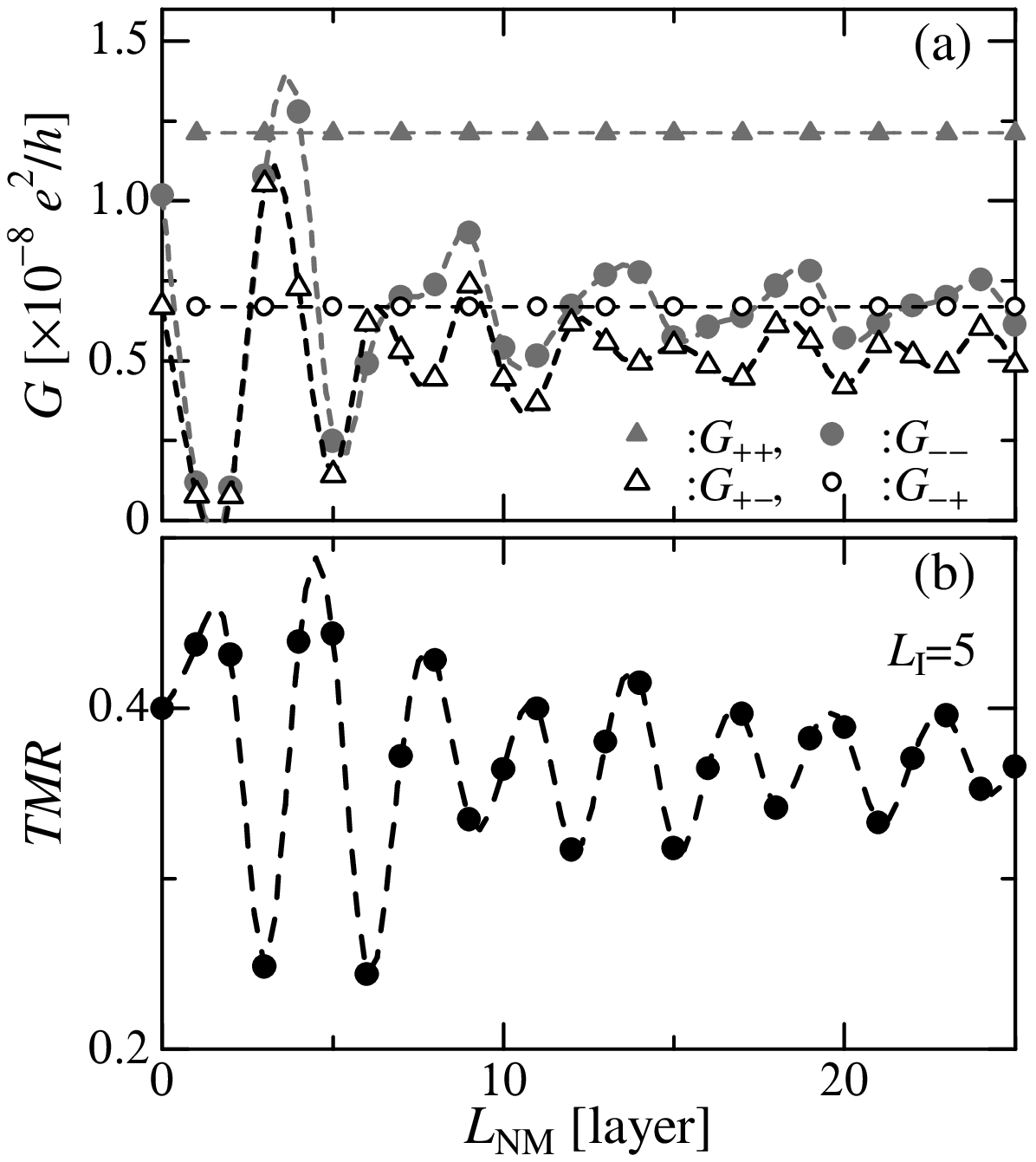}
\caption{Spin dependent conductance (a) and TMR ratio (b) calculated 
for clean junctions. 
Conductances for $\uparrow$ and $\downarrow$ spin electrons are 
plotted by triangles and circles, respectively, 
for parallel (solid symbols) and antiparallel alignment (open symbols) 
of magnetizations. 
Dashed lines are visual aids. }
\label{fig2}
\end{center}
\end{figure}

Figure 3 shows the dependence of oscillations in $G_{--}$ on 
the barrier thickness $L_{\mathrm{I}}$. 
Here the conductances are normalized to the asymptotic values 
$G_{--}^{\infty }$ obtained for $L_{\mathrm{NM}}\rightarrow \infty$. 
It can be seen that the oscillation period tends to $5a$ 
with increasing barrier thickness. 
This result is explained as follows. 
The wave vector $\mathbf{k}_{\parallel }$ parallel to the interface 
is conserved in the system without disorder, that is, 
$t_{\sigma \sigma ^{\prime }} 
(\mathbf{k}_{\parallel }\rightarrow \mathbf{k^{\prime }}_{\parallel }) 
= t_{\sigma \sigma ^{\prime }}(\mathbf{k}_{\parallel }) 
\delta _{\mathbf{k}_{\parallel },\mathbf{k^{\prime}}_{\parallel }}$. 
The transmission coefficient depends strongly on the  angle of incidence 
of electrons tunneling across the barrier, and the normal incidence 
contributes most to the conductance. 
It follows that, as the barrier thickness increases, 
the oscillation given by 
cut-off k-points, i.e., $\mathbf{k}_{\parallel }\neq \mathbf{0}$, 
becomes progressively weakened compared to that given by 
the Fermi wave vector of NM, i.e., $\mathbf{k}_{\parallel }=\mathbf{0}$. 
As for $G_{+-}$, we could not see the increase in oscillation period 
from $3a$ to $5a$ unless we increase $L_{\mathrm{I}}$ further. 
This might be due to the fact that 
$t_{+-}(\mathbf{k}_{\parallel }=\mathbf{0}) 
/\sum_{\mathbf{k}_{\parallel }}t_{+-}(\mathbf{k}_{\parallel })$ 
is smaller than 
$t_{--}(\mathbf{k}_{\parallel }=\mathbf{0}) 
/\sum_{\mathbf{k}_{\parallel}}t_{--}(\mathbf{k}_{\parallel })$: 
that is, the contribution of normal incidence to the conductance 
in $G_{+-}$ is less than that in $G_{--}$. 
As a result, 
the oscillation period of the TMR ratio is not quite $5a$ 
for the present barrier thickness. 

\begin{figure}
\begin{center}
\leavevmode
\includegraphics[width=0.65\linewidth,clip]{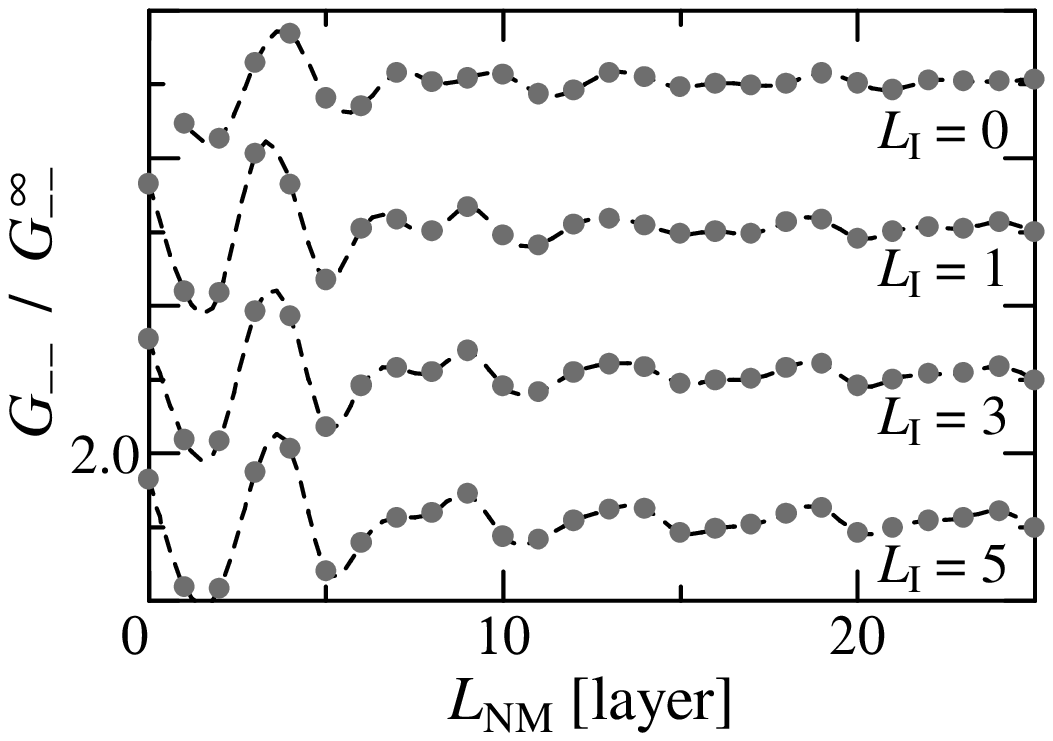}
\caption{Conductance $G_{--}$ for various thicknesses 
of clean insulating barrier. 
Conductances are normalized by the asymptotic value $G_{--}^\infty$ 
obtained within the limit of large spacer thickness. 
Dashed lines are visual aids. }
\label{fig3}
\end{center}
\end{figure}

We now introduce disorder into the insulating barrier 
and show the calculated results of the spin-dependent conductances 
and TMR ratios in Figs.~4(a) and 4(b), respectively. 
It can be seen that the conductance $G_{+-}$ is enhanced 
by disorder whereas the other conductances 
$G_{++}$, $G_{--}$, and $G_{-+}$ are hardly affected. 
$G_{+-}$ now oscillates almost exclusively with period $5a$, (i.e.
$k_{\mathrm{F}}$ period) about $G_{++}$ (see Fig. 4(a)). 
This results in a TMR ratio which is decreased and oscillates around zero 
with period $5a$. 
This should be contrasted with the ordered case 
in which the TMR ratio oscillates with a mixed period 
about a constant background (cf. Fig. 4(b) and Fig. 2(b)). 
The asymptotic values of the TMR ratio as 
$L_{\mathrm{NM}} \rightarrow \infty$ are shown 
in the inset of Fig.~4(b) 
as functions of the barrier thickness. 
Both the cases with and without disorder are shown. 
The asymptotic value of the TMR ratio of junctions without disorder 
decreases slowly with increasing $L_{\mathrm{I}}$, 
whereas that of junctions with disorder decreases rapidly and becomes 
zero for large $L_{\mathrm{I}}$. 

\begin{figure}
\begin{center}
\leavevmode
\includegraphics[width=0.70\linewidth,clip]{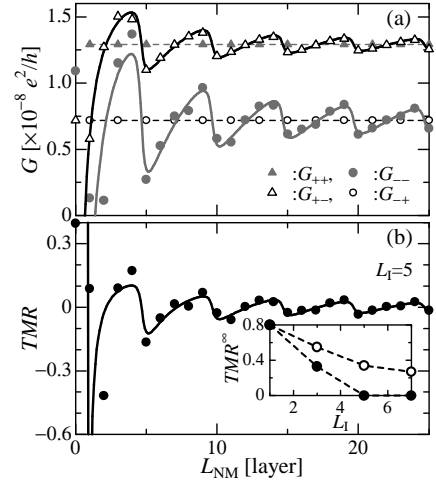}
\caption{Spin dependent conductance (a) and TMR ratio (b) calculated 
for disordered junctions. 
Conductances for $\uparrow$ and $\downarrow$ spin electrons are plotted 
by triangles and circles, respectively, 
for parallel (solid symbols) and antiparallel alignment (open symbols) 
of magnetizations. 
Solid lines indicate results obtained by the stationary phase approximation 
while dashed lines are visual aids. 
Inset of (b): Asymptotic values of TMR ratio $TMR^\infty$ 
obtained within the limit of large spacer thickness calculated 
for clean (open circles) and disordered (solid circles) junctions. }
\label{fig4}
\end{center}
\end{figure}

To gain a better understanding of the effects of disorder 
on the magnitude of $G_{+-}$, and on the period of oscillations, 
we have calculated the dependence of the transmission coefficient 
on $\mathbf{k}_{\parallel }$. 
Figures 5(a) and 5(b) show the transmission coefficients 
$T_{+-}(\mathbf{k}_{\parallel })$, 
where $T_{\sigma \sigma ^{\prime }}(\mathbf{k}_{\parallel }) \equiv 
\sum_{\mathbf{k^{\prime }}_{\parallel }} t_{\sigma \sigma ^{\prime }} 
(\mathbf{k}_{\parallel } \rightarrow \mathbf{k^{\prime }}_{\parallel })$ 
is the transmission coefficient 
for an electron incident from the left FM 
with momentum $\mathbf{k}_{\parallel }$ on the barrier 
without and with disorder, respectively. 
When there is no disorder, the contribution to $T_{+-}$ 
in the momentum space is concentrated near $\mathbf{k}_{\parallel }=(0,0)$.
However, inclusion of disorder gives rise to additional contributions 
to $T_{+-}$ of momenta outside this area. 
This is due to the fact that $\mathbf{k}_{\parallel}$ 
need not be conserved in diffusive scattering. 
In the absence of  disorder, only $\mathbf{k}_{\parallel}$ points 
on the Fermi surfaces, 
that satisfy the $\mathbf{k}_{\parallel}$ conservation, 
may contribute to the conductance. 
For diffusive scattering, on the other hand, 
the entire set of $\mathbf{k}_{\parallel}$  points on the Fermi surface 
contributes to the transport. 
More precisely, $\mathbf{k}_{\parallel }$ points 
corresponding to these quantum well states contribute to the conductance. 
These $\mathbf{k}_{\parallel}$ points appear as spiky peaks in Fig.~5(b) 
and they fall on 
concentric rings in the $\mathbf{k}_{\parallel}$ space.

\begin{figure}
\begin{center}
\leavevmode
\includegraphics[width=0.65\linewidth,clip]{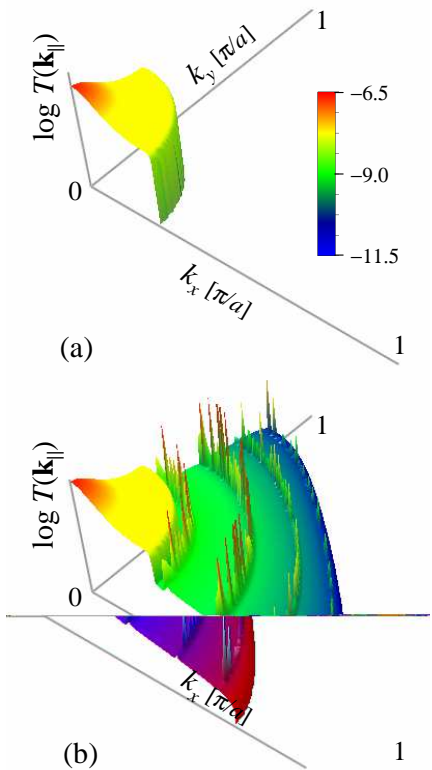}
\caption{$\mathbf{k}_\parallel$-dependence of transmission coefficients 
$T_{+-}$ calculated for (a) clean and (b) disordered junctions.}
\label{fig5}
\end{center}
\end{figure}

It is clear from Fig. 5(b) that the number of open $\mathbf{k}_{\parallel}$ 
channels contributing to $T_{+-}$, is the same as that contributing 
to $T_{++}$.
This explains the increase in the constant part of the conductance $G_{+-}$ 
to a value of approximately $G_{++}$.
In addition, the introduction of diffusive scattering 
has almost eliminated the sharp momentum cut-off observed in fig. 5(a), 
which explains why the $k_{\mathrm{cp}}$ oscillation period of $G_{+-}$ 
is weakened by disorder.
The other transmission coefficients are not greatly affected 
by the introduction of disorder, as scattering cannot open 
new $\mathbf{k}_{\parallel}$ channels for these cases. 
This explains why the introduction of disorder has little affect 
on the conductances $G_{++}$, $G_{--}$, and $G_{-+}$. 

We therefore expect that in the presence of disorder, 
the oscillatory part of the conductance is derived entirely from states 
in the region of $\mathbf{k}_{\parallel }=(0,0)$. 
In order to check this hypothesis, we use the stationary phase method 
(see Ref.\cite{mathon4} and references therein), 
which is able to determine the oscillatory contributions 
from isolated regions of the Brillouin zone, for thick spacers. 
The results, depicted by solid curves in fig.~4(a) and 4(b), 
are in excellent agreement with the numerical calculations 
for $L_{\mathrm{NM}} \gtrsim 5$. 
This fact indicates that the oscillation period observed 
in the experiments may be determined in the stationary phase approximation 
for realistic systems. 
The experimental finding \cite{yuasa2} 
that the oscillation period is determined by $k_{\mathrm{F}}$ of Cu spacer 
is thus naturally explained. 
Realistic calculation for the TMR oscillation and its bias dependence
is in progress. 

In summary, 
the period of oscillation determined by $k_{\mathrm{F}}$ of the spacer 
is dominant in TMR due to 
the $\mathbf{k}_{\parallel }$-selection by the insulating barrier 
and the deconfinement of the quantum well states by disorder. 
The diffusive scattering caused by the disorder increases 
the conductance in antiparallel alignment 
by opening new conductance channels via quantum well states 
and results in the oscillation of the TMR ratio 
around an averaged value close to zero. 
The success of the stationary phase approximation in reproducing the 
numerical results indicates that the oscillation period observed in
realistic tunneling junctions can be explained
in terms of $k_{\mathrm{F}}$ of Cu spacer.

\begin{acknowledgments}
H.I. would like to thank the visiting fellow program of the JSPS and the 
CREST Suzuki team for their valuable discussions and the financial support. 
J.I. acknowledges the financial support of the NEDO 
International Joint Research Project (NAME). 
\end{acknowledgments}


\end{document}